# Nonlinear compressed sensing based on composite mappings and its pointwise linearization

Jiawang Yi[1,2], Guanzheng Tan[1]

[1]School of Information Science and Engineering, Central South University, Changsha 410083, China.

[2]School of Computer & Communication Engineering, Changsha University of Science & Technology, Changsha 410114, China.

Emails: {JiawangYi, tgz}@csu.edu.cn

**Abstract:** Classical compressed sensing (CS) allows us to recover structured signals from far few linear measurements than traditionally prescribed, thereby efficiently decreasing sampling rates. However, if there exist nonlinearities in the measurements, is it still possible to recover sparse or structured signals from the nonlinear measurements? The research of nonlinear CS is devoted to answering this question. In this paper, unlike the existing research angles of nonlinear CS, we study it from the perspective of mapping decomposition, and propose a new concept, namely, nonlinear CS based on composite mappings. Through the analysis of two forms of a nonlinear composite mapping $\Phi$, i.e., $\Phi(x) = F(Ax)$ and $\Phi(x) = AF(x)$, we give the requirements respectively for the sensing matrix $A$ and the nonlinear mapping $F$ when reconstructing all sparse signals exactly from the nonlinear measurements $\Phi(x)$. Besides, we also provide a special pointwise linearization method, which can turn the nonlinear composite mapping $\Phi$, at each point in its domain, into an equivalent linear composite mapping. This linearization method can guarantee the exact recovery of all given sparse signals even if $\Phi$ is not an injection for all sparse signals. It may help us build an algorithm framework for the composite nonlinear CS in which we can take full advantage of the existing recovery algorithms belonging to linear CS.

**Keywords**：Compressed sensing, Nonlinear measurements, Composite mappings, Restricted isometry property, Invariance, Pointwise linearization.

## 1 Introduction

Compressed sensing (CS) [4, 7, 10] allows us to recover sparse or structured signals from few measurements. Specifically, assume that we sample a $k$-sparse signal $x \in R^n$ using $m$ linear measurements ($m < n$) and collect the measurements into a vector $y \in R^m$, then the sampling process can be described as $y = Ax$, where $A$ is an $m \times n$ sensing matrix or measurement matrix. When all $k$-sparse signals have different projections under $A$, naturally $x$ becomes the sparsest one of those vectors consistent with the measurements $y$. Therefore, we can recover $x$ by solving an optimization problem of the form

$$\hat{x} = \arg\min_{u} \|u\|_0 \ \ subject\ to\ \ y = Au\ . \tag{1.1}$$

As is known to all, (1.1) is an NP-hard problem [17] for the object function $\|\cdot\|_0$ is not convex. We may replace $\|\cdot\|_0$ with its convex approximation $\|\cdot\|_1$, and then have a relaxed and more tractable optimization problem

$$\hat{x} = \arg\min_{u} \|u\|_1 \quad subject\ to \quad y = Au. \tag{1.2}$$

We know that nonlinear observations or samples exist extensively in the real world and many of them are inevitable. Hence, it is necessary to extend CS to nonlinear models, i.e., nonlinear CS [1, 2, 23]. Its mainest difference from classical linear CS lies in replacing the linear mapping, i.e., the matrix $A$, with a nonlinear mapping $\Phi$. Via this replacement, respectively from (1.1) and (1.2) we can get the $\ell_0$ minimization problem of nonlinear CS,

$$\hat{x} = \arg\min_{u} \|u\|_0 \quad subject\ to \quad y = \Phi(u), \tag{1.3}$$

and its relaxed version, i.e., $\ell_1$ minimization,

$$\hat{x} = \arg\min_{u} \|u\|_1 \quad subject\ to \quad y = \Phi(u). \tag{1.4}$$

Like linear CS, dimensionality reduction is still one of the necessary characteristics possessed by nonlinear CS. It means that we have $m < n$ in $y = \Phi(x)$, where $m$ and $n$ are the dimensionalities of the nonlinear measurements $y$ and the sparse signal $x$ respectively.

So far, the literature on nonlinear CS is still very limited. The research of nonlinear CS was first seen in the work about compressive sensing phase retrieval (CSPR) by Moravec, Romberg, and Baraniuk [16], and the work about 1-bit CS by Boufounos and Baraniuk [3]. These authors tried the reconstruction of sparse signals in these two nonlinear CS examples by employing the knowledge of CS. In [2], Blumensath gave the theoretical generalization of nonlinear CS for the first time, and showed that, under conditions similar to the restricted isometry property (RIP), the iterative hard thresholding (IHT) algorithm can be used to accurately recover sparse or structured signals from few nonlinear measurements. The research of quadratic basis pursuit or nonlinear basis pursuit by Ohlsson *et al.* [18-21], and the research of sparsity constrained nonlinear optimization by Beck and Eldar [1], also contribute to the literature on nonlinear CS.

In this paper, on the basis of analyzing nonlinear CS examples, we decompose $\Phi$ into a CS mapping and a nonlinear mapping, study nonlinear CS from this new angle of view, and therefore propose a new concept, nonlinear CS based on composite mappings. Apparently, one cannot recover all $k$-sparse signals (collectively denoted by $\Sigma_k$) exactly based merely on the measurements unless the nonlinear composite mapping $\Phi$ is injective for all $x \in \Sigma_k$. Under this condition, we analyze the sparse recovery problem of two forms of the composite mapping $\Phi$, i.e., $\Phi(x) = F(Ax)$ and $\Phi(x) = AF(x)$. Meanwhile, based on our research result in [24], which is about the invariance of sparse recovery properties under matrix elementary transformations, we also propose a pointwise linearization method that can replace the nonlinear composite mapping $\Phi$, at each point in its domain, with an equivalent linear composite mapping of certain types. Moreover, these special linear composite mappings all satisfy the RIP of order $2k$, which guarantees the accurate reconstruction of all given $k$-sparse signals from the nonlinear measurements $\Phi(x)$.

The remainder of this paper is organized as follows. In Section 2, we introduce sparse recovery properties such as the spark, null space property (NSP) and RIP, and our research finding about the invariance of these three properties under matrix elementary transformations. In Section 3, we study the sparse recovery problem of the nonlinear CS based on composite mappings. First, through the analysis of nonlinear CS examples, we propose an idea that redefines the nonlinear mapping $F$ as, or turns it into, a pointwise linear mapping, and provide the conditions of the pointwise equivalent replacement of $F$ with certain type matrices $M_I$ and $M_D$. Second, as for two forms of the nonlinear composite mapping Φ, i.e., Φ(*x*) = *F*(*Ax*) and Φ(*x*) = *AF*(*x*), we analyze and discuss the requirements of exact sparse recovery for *F* and *A*. Then, we give a special pointwise linearization method that can perform the pointwise transformation of the nonlinear CS based on composite mappings into linear CS, and make us able to solve the signal recovery problem of nonlinear CS with the recovery algorithms of linear CS. In Section 4, we take several experiments as examples to demonstrate the capability of this method to reconstruct sparse signals accurately from nonlinear measurements. In Section 5, we conclude with a discussion on some open questions as well as future research work.

## 2 Background

### 2.1 Sparse recovery properties

When a sensing matrix *A* satisfies certain conditions related to the following three sparse recovery properties, it can be guaranteed that all *k*-sparse signals have no overlapped projections under *A*. One of the most fundamental properties is the spark [11].

**Definition 2.1.** *The spark of a given matrix A is the smallest number of columns of A that are linearly dependent.*

If spark(*A*) > 2*k*, there exist no two signals $x, x' \in \Sigma_k$ such that $Ax = Ax'$. The converse is also true.

The null space property (NSP) [8] enables the error of reconstructing a sparse or compressible signal to be measured via that of its best *k*-term approximation. When *A* satisfies the NSP of order 2*k*, one can recover all *k*-sparse signals exactly from noise-free measurements, for their best *k*-term approximation errors are all zero.

**Definition 2.2.** *A matrix A satisfies the NSP of order k if there exists a constant C > 0 such that*

$$\|h_\Lambda\|_2 \leq C \frac{\|h_{\Lambda^c}\|_1}{\sqrt{k}} \tag{2.1}$$

*holds for all* $h \in \mathcal{N}(A)$ *and for all* Λ *such that* |Λ| ≤ *k*.

The restricted isometry property (RIP) [5] can not only contain the NSP but also make *A* able to approximately preserve the distance between any pair of sparse signals. This feature promotes the robustness to noise of the RIP.

**Definition 2.3.** *A matrix A satisfies the RIP of order k if there exists a* $\delta_k \in (0, 1)$ *such that*

$$(1-\delta_k)\|x\|_2^2 \leq \|Ax\|_2^2 \leq (1+\delta_k)\|x\|_2^2 \tag{2.2}$$

*holds for all* $x \in \Sigma_k$.

Actually, the bounds in Definition 2.3 of the RIP need not to be symmetric about 1. An asymmetric version of the RIP definition [9] is given below.

**Definition 2.4.** *A matrix A satisfies the RIP of order k if there exist constants $\alpha$ and $\beta$ ($0 < \alpha \leq \beta < \infty$) such that*

$$\alpha \|x\|_2^2 \leq \|Ax\|_2^2 \leq \beta \|x\|_2^2 \tag{2.3}$$

*holds for all $x \in \Sigma_k$.*

It is easy for us to convert the RIP from the asymmetric version into the symmetric version. When $A$ satisfies the asymmetric RIP of order $k$, one can multiply $A$ by $\lambda = \sqrt{2/(\beta+\alpha)}$ and then the product $\lambda A$ satisfies the symmetric RIP of order $k$ with constant $\delta_k = (\beta - \alpha)/(\beta + \alpha)$.

Candès [6] has shown that, when $A$ satisfies the symmetric RIP of order $2k$, we have that: if $\delta_{2k} < 1$ then $\ell_0$ minimization problem (1.1) has a unique $k$-sparse solution; if $\delta_{2k} < \sqrt{2} - 1$ then $\ell_1$ minimization problem (1.2) has the same solution as that of $\ell_0$ minimization problem (1.1).

**2.2 Invariance of sparse recovery properties under matrix elementary transformations**

Before this paper, we studied, under matrix elementary transformations, the invariance of the following three properties of sparse recovery, i.e., the spark, NSP and RIP. We found that the spark, NSP order and RIP order of a sensing matrix $A$ will not be changed if one apply to $A$ all elementary row or column operations except column addition. For the proof and details please refer to [24]. In linear algebra, applying to $A$ elementary row or column operations equals to multiplying $A$ by their corresponding elementary matrices. Hence, it is easy to derive the following Theorems 2.1-2.3 (Corollaries 3.1-3.3 in [24]). These theorems indicate that the products of the sensing matrix $A$ and certain types of matrices can preserve the above three property values. This invariance will be used later when we consider the pointwise linearization of nonlinear mappings.

**Theorem 2.1.** *Let $B_1$ be an invertible matrix, and $B_2$ be a matrix produced by any permutation of the rows or columns of an invertible diagonal matrix. For a matrix A, we have $\text{spark}(B_1A) = \text{spark}(A)$ and $\text{spark}(AB_2) = \text{spark}(A)$.*

**Theorem 2.2.** *Let $B_1$ be an invertible matrix, and $B_2$ be a matrix produced by any permutation of the rows or columns of an invertible diagonal matrix. If a matrix A satisfies the NSP of order k, then $B_1A$ and $AB_2$ still satisfy the NSP of order k.*

**Theorem 2.3.** *Let $B_1$ be an invertible matrix, and $B_2$ be a matrix produced by any permutation of the rows or columns of an invertible diagonal matrix. If a matrix A satisfies the asymmetric RIP of order k, then $B_1A$ and $AB_2$ still satisfy the asymmetric RIP of order k.*

**3 Nonlinear CS based on composite mappings**

**3.1 Examples of nonlinear CS**

We provide below several practical examples that can be classified into the nonlinear CS based on composite mappings.

Phase retrieval (PR) is such a problem that we can only observe the magnitude (or intensity) of a signal's Fourier transform, but want to recover its phase and therefore the signal itself [12, 13]. If we randomly sub-sampled the magnitude data and render the number of the measurements smaller than the dimensionality of the signal, the effect of doing so is equivalent to taking magnitude after multiplying the signal $x$ by a CS sensing matrix $A$ that is formed by randomly selected rows from the discrete Fourier transform matrix [4, 7, 10]. It gets us finally the magnitudes of the CS measurements $|Ax|$, where $|\cdot|$ denotes taking the magnitude (absolute value or modulus) of each element. The process of recovering a signal from this kind of nonlinear CS measurements is referred to as compressive sensing phase retrieval (CSPR) [16].

When computers process CS measurements $Ax$, the measurements must be mapped from real values into discrete values represented by finite bits, which is called quantization. The quantization process can be regarded as applying a truncation function to the CS measurements, which results in trunc($Ax$), where trunc($\cdot$) denotes the element-wise truncation. Quantization inevitably introduces error in the CS measurements and this quantization error is generally treated as bounded noise. In an extreme quantization case, we use a single bit to represent the sign of each CS measurement, i.e., sign($Ax$), where the signum function sign($\cdot$) is applied element-wise to $Ax$. The process of recovering a signal from sign($Ax$), a type of nonlinear CS measurements, is referred to as 1-bit CS [3, 15, 22].

The output of an ideal sensor is supposed to always represent the true values of external stimuli. That is to say, when the ideal sensor senses true signals $x$, its output is also $x$. However, in practice, real transfer functions of sensors are generally nonlinear [14]. The error brought by this nonlinearity may grow bigger when sensors are subjected to factors such as aging. In the case that this nonlinearity error needs to be considered or can not be ignored, the output of a sensor will no longer be viewed as true signals $x$ but their nonlinear function values $F(x)$. Here we assume that, when there is no stimulus, we can adjust the output of the sensor to zero, i.e., $F(\mathbf{0}) = \mathbf{0}$. And in this paper we only consider the nonlinear sensors satisfying this assumption. So, for sensors in CS sampling systems, their nonlinearity of this sort will turn CS measurements $Ax$ into nonlinear CS measurements $AF(x)$.

**3.2 Analysis of nonlinear CS examples**

In the above examples of nonlinear CS, the finally observed nonlinear measurements $\Phi(x)$ can be regarded as two forms of composite mappings, i.e., $\Phi = F \circ L_A$ or $\Phi = L_A \circ F$, where $L_A$ is a linear mapping of classical CS with the sensing matrix $A$ as its representation, and $F$ is a nonlinear function or mapping. In the analysis of these nonlinear CS examples, we used vector-valued functions to represent the nonlinear mappings $F$ and made some appropriate treatment, then we found that $F$ can be equivalent, at each point in its domain, to an invertible diagonal matrix (probably more than one).

(1) *the first composition form* $\Phi = F \circ L_A$

This type of nonlinear measurements is formed by applying nonlinear mappings after CS. The corresponding examples are CSPR in the case of $R^n$ and quantization of CS measurements including 1-bit CS.

If we represent $F$ with a vector-valued function, then we have

$$\Phi(\boldsymbol{x}) = F(A\boldsymbol{x}) = F(\boldsymbol{y}) = (f_1(y_1), f_2(y_2), \ldots, f_m(y_m))^T.$$

In the corresponding examples, all component functions of $F$ are the same, i.e., $f_i = f$ ($i = 1, 2, \ldots, m$), and $f$ can be functions of absolute value, truncation and signum. We know that these functions all satisfy that $f(y_i) = 0$ if and only if $y_i = 0$. This means that: when $y_i \neq 0$, there must exist some real number $c_i \neq 0$ such that $f(y_i) = c_i y_i$; and when $y_i = 0$, because $f(y_i) = 0$ any real number $c_i \neq 0$ satisfies that $f(y_i) = c_i y_i$. Hence, for any given measurement vector $\boldsymbol{y} \in \text{dom } F$, there must exist an invertible diagonal matrix (probably not unique) such that

$$F(\boldsymbol{y}) = (c_1 y_1, c_2 y_2, \ldots, c_m y_m)^T = \text{diag}(c_1, c_2, \ldots, c_m)(y_1, y_2, \ldots, y_m)^T, \qquad (3.1)$$

where all diagonal entries $c_i \neq 0$. Here we must point out that (3.1) holds only when $\boldsymbol{y}$ is given or fixed. Finally, according to (3.1), we see that the first composition form is transformed, at each point in dom $F$, into a composition of the CS mapping with a linear mapping, i.e., $F(A\boldsymbol{x}) = \text{diag}(c_1, c_2, \ldots, c_m)A\boldsymbol{x}$.

(2) *the second composition form* $\Phi = L_A \circ F$

This type of nonlinear measurements is formed by applying nonlinear mappings before CS. The corresponding example is nonlinear CS sensors. In CS sampling systems based on distributed sensor networks, imaging sensors, etc., the sensed true signals $\boldsymbol{x} \in R^n$ consist directly of the outputs of $n$ sampling sensors. But when the sensors possess considerable nonlinearity, the outputs needs to be taken as $F(\boldsymbol{x})$ and what the CS systems observe is this form of nonlinear measurements.

Likewise, we use a vector-valued function to represent $F$ and get

$$\Phi(\boldsymbol{x}) = AF(\boldsymbol{x}) = A(f_1(x_1), f_2(x_2), \ldots, f_n(x_n))^T.$$

In the corresponding example, different sensors typically have different $f_i$ ($i = 1, 2, \ldots, n$). If the outputs of these CS sensors satisfy the assumption that $F(\boldsymbol{0}) = \boldsymbol{0}$, these nonlinear or linear $f_i$ all satisfy that $f_i(x_i) = 0$ if and only if $x_i = 0$. This means that: when $x_i \neq 0$, there must exist some real number $c_i \neq 0$ such that $f_i(x_i) = c_i x_i$; and when $x_i = 0$, because $f_i(x_i) = 0$ any real number $c_i \neq 0$ satisfies that $f_i(x_i) = c_i x_i$. Thus, for any given true signal $\boldsymbol{x} \in \text{dom } F$, there must exist an invertible diagonal matrix (probably not unique) such that

$$F(\boldsymbol{x}) = (c_1 x_1, c_2 x_2, \ldots, c_n x_n)^T = \text{diag}(c_1, c_2, \ldots, c_n)(x_1, x_2, \ldots, x_n)^T, \qquad (3.2)$$

where all diagonal entries $c_i \neq 0$. Similarly, (3.2) holds only for the given or fixed $\boldsymbol{x}$. At last, based on (3.2), we can transform the second composition form, at each point in dom $F$, into a composition of the CS mapping with a linear mapping, i.e., $AF(\boldsymbol{x}) = A\text{diag}(c_1, c_2, \ldots, c_n)\boldsymbol{x}$.

In fact, initially we did not know to correlate $F$ with an invertible diagonal matrix at each point in dom $F$. In analyzing the nonlinear CS examples, we just found that these nonlinear mappings $F$ can be represented, pointwise, with a diagonal matrix (linear mapping) after appropriate representation and treatment of $F$. We were inspired from this to pose the question: what kind of diagonal matrices, when one multiplies a sensing matrix $A$ by them, can preserve the sparse recovery capability of $A$? For sparse recovery properties, such as the spark, NSP, and RIP, can characterize this capability of $A$, we studied the invariance of these properties under matrix elementary transformations, and thus learned the types of those matrices whose products with $A$ can preserve the spark, NSP order and RIP order of $A$. One of these types is invertible diagonal matrices. So we began to realize that the nonlinear mappings $F$ in the examples can actually be equivalent, pointwise, to an invertible diagonal matrix.

**3.3 Pointwise linearization of nonlinear mappings**

From the analysis of the nonlinear CS examples, we realize that the nonlinear mappings $F$ can be equivalent, at each point in $F$'s domain, to a linear mapping (matrix), and the corresponding linear mappings at distinct points may or can vary. We know that a piecewise function in mathematics can define different expressions in different intervals of its domain, thereby introducing concepts such as piecewise linear, piecewise continuous, piecewise differentiable, etc. In an extreme case, we can restrict every interval of a piecewise function's domain to a point. If we redefine $F$ as a piecewise function of this extreme form, as above described, $F$ can be redefined, at each point in its domain, as an equivalent linear mapping. We may refer to the redefined $F$ as a pointwise linear mapping or function. "Pointwise linear" means that $F$ is nonlinear for the whole domain but linear for its each point. According to Definition 3.1 below, we can redefine a nonlinear mapping $F$ as, or turn it into, a pointwise linear mapping.

**Definition 3.1.** *Let F be a nonlinear mapping and* $\text{dom } F = \{z_1, z_2, z_3, \ldots\}$. *If, for any given* $z \in \text{dom } F$, *there always exists a matrix M such that* $F(z) = Mz$, *i.e.,*

$$F(z) = \begin{cases} M_1 z_1, & \text{if } z = z_1 \\ M_2 z_2, & \text{if } z = z_2 \\ M_3 z_3 & \text{if } z = z_3 \\ \vdots & \end{cases},$$

*then we call F a pointwise linear mapping.*

Here we point out that, the difference between the definition of a linear mapping and that of a pointwise linear mapping lies in that, the former requires that there exist a single matrix $M$ such that $F(z) = Mz$ for all $z \in \text{dom } F$.

In Section 3.2, we already know that in those composite mappings $\Phi = F \circ L_A$ or $\Phi = L_A \circ F$, the nonlinear mappings $F$ can be replaced, at any given point in $F$'s domain, with an equivalent invertible diagonal matrix. According to Theorems 2.1-2.3, the product of a sensing matrix $A$ with an invertible diagonal matrix possesses the same spark, NSP order and RIP order as $A$. This means that, not only is $F$ a pointwise linear mapping but also preserves these property values of $A$ at every point. Hence, based on Theorems 2.1-2.3, the following Definition 3.2 can guarantee that, in the composition of a nonlinear mapping $F$ with a linear mapping $L_A$ (sensing matrix $A$), $F$ becomes a pointwise linear mapping such that these property values of $A$ are preserved.

**Definition 3.2.** *Let* $L_A$ *be the corresponding linear mapping to a sensing matrix A, F be a nonlinear mapping. Also let* $M_I$ *be an invertible matrix, and* $M_D$ *be a matrix produced by any permutation of the rows or columns of an invertible diagonal matrix. We call F a pointwise linear mapping that preserves the spark, NSP order and RIP order of A, if F satisfies one of the following conditions:*

*(i) In the composite mapping* $F \circ L_A$, *for any given* $z \in \text{dom } F$, *there always exists a matrix* $M_I$ *such that* $F(z) = M_I z$.

*(ii) In the composite mapping* $L_A \circ F$, *for any given* $z \in \text{dom } F$, *there always exists a matrix* $M_D$ *such that* $F(z) = M_D z$.

For the composite mappings $\Phi = F \circ L_A$ or $\Phi = L_A \circ F$, if $F$ satisfies Definition 3.2, it means that $F$ can be equivalent, pointwise, to a matrix $M_I$ or $M_D$, and the corresponding matrix product $M_I A$ or $A M_D$ to $\Phi$ at each point possesses the same spark, NSP order and RIP order as $A$.

### 3.4 Requirements of pointwise linearization for nonlinear mappings

In Definition 3.2, it is required that the nonlinear mappings $F$ in those composite mappings be equivalent pointwise to a special matrix $M_I$ or $M_D$. So, what conditions $F$ must satisfy to guarantee that $F$ becomes one of these two types of special pointwise linear mappings? In other words, under what conditions we are able to apply to $F$ these two types of special pointwise linearization? For a stepwise deduction, we first derive the basic conditon of becoming a pointwise linear mapping, then arrive at the conditions of transforming $F$ into a pointwise linear mapping correlated to $M_I$ or $M_D$.

(1) *The requirement for $F$ of the first type of pointwise linearization, i.e., that "for any given $z \in$ dom $F$, there always exists a matrix $Y$ such that $F(z) = Yz$".*

This is the basic condition for a nonlinear mapping $F$ to become a pointwise linear mapping. From the angle of solving systems of equations, this type of pointwise linearization requires that there always be a solution for the matrix $Y$ in the following system of equations:

$$F(z) = \begin{pmatrix} f_1(z) \\ f_2(z) \\ \vdots \\ f_n(z) \end{pmatrix} = \begin{pmatrix} y_{11} & y_{12} & \cdots & y_{1n} \\ y_{21} & y_{22} & \cdots & y_{2n} \\ \vdots & \vdots & & \vdots \\ y_{n1} & y_{n2} & \cdots & y_{nn} \end{pmatrix} \begin{pmatrix} z_1 \\ z_2 \\ \vdots \\ z_n \end{pmatrix} = Yz. \qquad (3.3)$$

When $z$ is given or known, $F(z)$ or $f_i(z)$ ($i = 1, 2, \ldots, n$) is also fixed or known, and thus the unknowns in the equations $F(z) = Yz$ are just $y_{ij}$ ($i, j = 1, 2, \ldots, n$), whose number is $n^2$ in all. Note that the unknowns in the first row of $Y$ appear only in the first equation, i.e.,

$$(y_{11}, y_{12}, \ldots, y_{1n})(z_1, z_2, \ldots, z_n)^T = f_1(z).$$

According to the theory of solving inhomogeneous linear equations, we have the following conclusion: these unknowns always have solutions when $z \neq \mathbf{0}$, and also do, when $z = \mathbf{0}$, if and only if $f_1(\mathbf{0}) = 0$. Likewise, a similar conclusion holds, as well, for the unknowns in any other row of $Y$. Consequently, the matrix $Y$ always has solutions when $z \neq \mathbf{0}$, and also does, when $z = \mathbf{0}$, if and only if $F(\mathbf{0}) = \mathbf{0}$.

Hence, the first type of pointwise linearization requires $F$ to satisfy that "if $\mathbf{0} \in$ dom $F$ then $F(\mathbf{0}) = \mathbf{0}$".

(2) *The requirement for $F$ of the second type of pointwise linearization, i.e., that "for any given $z \in$ dom $F$, there always exists an invertible matrix $Y$ such that $F(z) = Yz$".*

For any given nonzero vector $z \in$ dom $F$, according to (1), the matrix $Y$ in the equations $F(z) = Yz$ always has solutions. If the matrix $Y$ is invertible, which means that its columns are linearly independent, then $F(z)$ cannot be $\mathbf{0}$. This is a necessary condition, and we will prove below that it is a sufficient condition too. That is to say, to prove that when the condition $F(z) \neq \mathbf{0}$ holds, we can always find or construct an invertible matrix $Y$ such that $F(z) = Yz$.

In (3.3), we note that the unknowns in the $i$ th row of $Y$ appear only in the $i$ th equation of the equations $F(z) = Yz$, i.e.,

$$(y_{i1}, y_{i2}, \ldots, y_{in})(z_1, z_2, \ldots, z_n)^T = f_i(z).$$

It means that, this inhomogeneous linear equation is the only constraint to be considered when determining the values of the unknowns of the $i$ th row of $Y$. Because $F(z) \neq \mathbf{0}$ and $z \neq \mathbf{0}$, we may assume that $f_p(z) \neq 0$ and $z_q \neq 0$ ($1 \leq p, q \leq n$). Then, we construct an invertible matrix $Y$, row by row, in the following way: (a) For the unknowns of row $p$ of $Y$, we set all of the unknowns except $y_{pq}$ to zero and thus have a nonzero element $y_{pq} = f_p(z) / z_q$. (b) For the unknowns of row $q$ of $Y$, let $y_{qp} = 1$ and $y_{qq} = (f_q(z) - y_{qp} z_p) / z_q$, and set the rest of these unknowns all to zero. (c) For the unknowns of row $i$ of $Y$ ($i = 1, 2, \ldots, n$ and $i \neq p, q$), let the diagonal entry $y_{ii} = 1$ and the column $q$ entry $y_{iq} = (f_i(z) - y_{ii} z_i) / z_q$, and set the rest of these unknowns all to zero. Now, all of the unknowns in the matrix $Y$ are determined. Take for example the case that $n = 5$, $p = 2$, and $q = 4$, the matrix $Y$ constructed by the above method is given as follows:

$$F(z) = \begin{pmatrix} f_1(z) \\ f_2(z) \\ f_3(z) \\ f_4(z) \\ f_5(z) \end{pmatrix} = \begin{pmatrix} 1 & 0 & 0 & y_{14} & 0 \\ 0 & 0 & 0 & y_{24} & 0 \\ 0 & 0 & 1 & y_{34} & 0 \\ 0 & 1 & 0 & y_{44} & 0 \\ 0 & 0 & 0 & y_{54} & 1 \end{pmatrix} \begin{pmatrix} z_1 \\ z_2 \\ z_3 \\ z_4 \\ z_5 \end{pmatrix} = Yz.$$

In the matrix $Y$ constructed in this way, one can exchange row $p$ and row $q$, and then convert all of the column $q$ entries except the nonzero entry $y_{pq}$ to zero by executing row addition operations with $y_{pq}$. Thus, we get a diagonal matrix whose rank is obviously $n$, and hence it is proved that $Y$ is an invertible matrix.

On the other hand, for a given zero vector $\mathbf{0} \in \text{dom } F$, there exists an invertible matrix $Y$ such that $F(\mathbf{0}) = Y \mathbf{0}$ if and only if $F(\mathbf{0}) = \mathbf{0}$. Therefore, the second type of pointwise linearization requires $F$ to satisfy that "if nonzero vectors $z \in \text{dom } F$ then $F(z) \neq \mathbf{0}$; and if $\mathbf{0} \in \text{dom } F$ then $F(\mathbf{0}) = \mathbf{0}$".

(3) *The requirement for F of the third type of pointwise linearization, i.e., that "for any given $z \in$ dom F, there always exists an invertible diagonal matrix Y such that $F(z) = Yz$".*

Let the matrix $Y$ be $\text{diag}(y_{11}, y_{22}, \ldots, y_{nn})$ and $y_{ii} \neq 0$ ($i = 1, 2, \ldots, n$). In this case, the number of the unknowns of the equations $F(z) = Yz$ is just $n$, so $Y$ does not always have solutions. From (3.3) we can get $f_i(z) = y_{ii} z_i$. Clearly, $y_{ii}$ has nonzero solutions if and only if $f_i(z)$ and $z_i$ become zero simultaneously. Therefore, the third type of pointwise linearization requires $F$ to satisfy that "$f_i(z) = 0$ if and only if $z_i = 0$ ($i = 1, 2, \ldots, n$)". Assuming that we take the validation of a necessary and sufficient condition as a basic operation, the time complexity for an algorithm to verify if $F$ satisfies this requirement is $O(n)$.

(4) *The requirement for F of the fourth type of pointwise linearization, i.e., that "for any given $z \in$ dom F, there always exists a matrix Y such that $F(z) = Yz$ and Y is produced by any permutation of the rows or columns of an invertible diagonal matrix".*

We know that there must exist some column exchange operations or their corresponding elementary matrices $Q_1, Q_2, \ldots, Q_t$ to transform $Y$ into an invertible diagonal matrix $Y^*$, and we have

$$F(z) = Yz = YQ_1 Q_2 \cdots Q_t Q_t^{-1} \cdots Q_2^{-1} Q_1^{-1} z = Y^* z^*,$$

where multiplying $z$ by the inverse matrix product $Q_t^{-1} \cdots Q_2^{-1} Q_1^{-1}$ on the left equals to transforming $z$ into $z^*$ by applying the corresponding row exchange operations. Based on (3), we know that the fourth type of pointwise linearization requires $F$ to satisfy that "there exists some permutation of all entries of $z$, denoted by $z^*$, such that $f_i(z) = 0$ if and only if $z_i^* = 0$ ($i = 1, 2, \ldots, n$)". Using the same basic operation as (3), the time complexity for an algorithm to verify if $F$ satisfies this requirement is $O(n)$ at best and $O(n \cdot n!)$ at worst.

From Section 3.2 we know that, in those nonlinear CS examples, $F$ can all be transformed into a pointwise linear mapping correlated to invertible diagonal matrices. These nonlinear mappings $F$ all satisfy that "$f_i(z) = f_i(z_i) = 0$ if and only if $z_i = 0$". It conforms to the requirement of the third type of pointwise linearization for $F$.

### 3.5 Recovery of sparse signals

In this section, we study the sparse recovery problem of the nonlinear CS based on composite mappings ($\Phi = F \circ L_A$ or $\Phi = L_A \circ F$). First, for the goal of exactly recovering all sparse signals, we will analyze and derive the recovery conditions to be satisfied respectively by $\Phi$, $F$ and $A$. Second, we will introduce a special linearization method, i.e., pointwise linearization method. When distinct sparse signals have the same nonlinear measurements, this method can still guarantee the accurate reconstruction of them.

#### 3.5.1 Analysis of recovery conditions

In nonlinear CS, for the true signals $x$ to be recovered are $k$-sparse, naturally we also hope to recover $x$, as in linear CS, via the associated $\ell_0$ minimization (1.3) or its relaxed version, namely $\ell_1$ minimization (1.4). It is easy to know that, based solely on measurements, one can exactly recover all $k$-sparse signals via (1.3) if and only if $\Phi$ is injective for all $k$-sparse signals. This condition is sufficient, because it can make any $x \in \Sigma_k$ become the sparsest one of those signals consistent with the projection $y = \Phi(x)$. We can guarantee this condition by prescribing a condition similar to the RIP to be satisfied by $\Phi$. For example, there exists a constant $0 < \delta < 1$ such that

$$(1-\delta)\|x_1 - x_2\|_2^2 \leq \|\Phi(x_1) - \Phi(x_2)\|_2^2 \leq (1+\delta)\|x_1 - x_2\|_2^2$$

holds for all $x_1, x_2 \in \Sigma_k$.

The nonlinear CS we consider here is based on composite mappings, hence in (1.3) the composite mappings $\Phi(x) = F(Ax)$ or $\Phi(x) = AF(x)$. Below we analyze the sparse recovery problem of this kind of nonlinear CS, by means of the decomposition of $\Phi$ as $F \circ L_A$ or $L_A \circ F$.

(1) $\Phi(x) = F(Ax)$

(a) When $A$ satisfies "spark > $2k$" (or the NSP or RIP of order $2k$), in order to ensure that $\Phi$ is an injection for all $x \in \Sigma_k$, it is required that $F$ must be injective for all linear projections $Ax$ ($x \in \Sigma_k$). Apparently, setting that $F$ is injective on dom $F$ is enough to satisfy this requirement. Or we may prescribe some weaker condition similar to the RIP to meet the requirement. For example, there exists a constant $0 < \delta < 1$ such that

$$(1-\delta)\|Ax_1 - Ax_2\|_2^2 \leq \|F(Ax_1) - F(Ax_2)\|_2^2 \leq (1+\delta)\|Ax_1 - Ax_2\|_2^2$$

holds for all $x_1, x_2 \in \Sigma_k$, where $A$ satisfies "spark > $2k$".

(b) When $A$ does not satisfy "spark > $2k$", there must exist (probably do if not satisfy the NSP or RIP of order $2k$) two distinct $k$-sparse signals $x_1$ and $x_2$ such that $Ax_1 = Ax_2 = z$. Because the image of $z$ under the mapping or function $F$ is unique, we have $F(Ax_1) = F(Ax_2)$. It means that $\Phi$ is not injective for all $k$-sparse signals.

(2) $\Phi(x) = AF(x)$

(a) When $F$ is injective for all $x \in \Sigma_k$, assume that, for all vectors $F(x)$ ($x \in \Sigma_k$), the maximum number of nonzero entries of a vector is $l$. We know that the spark of an $m \times n$ matrix $A$ is at most $m + 1$. Thus, if $l \leq m/2$ (meaning that these $F(x)$ are all $l$-sparse vectors), $A$'s satisfying "spark > $2l$" (or the NSP or RIP of order $2l$) is enough to ensure that $A$ is injective for all of the nonlinear projections $F(x)$ ($x \in \Sigma_k$), thereby ensuring that $\Phi$ is injective for all $k$-sparse signals. If $l > m/2$, there exists no $A$ able to satisfy "spark > $2l$". In this case, it is still possible that $A$ is injective for all $F(x)$ ($x \in \Sigma_k$). As in (1), no matter $l > m/2$ or $l \leq m/2$, we can also prescribe some condition similar to the RIP to ensure that $A$ is this kind of injection, thereby establishing that $\Phi$ is injective for all $x \in \Sigma_k$. For example, there exists a constant $0 < \delta < 1$ such that

$$(1-\delta)\|F(x_1) - F(x_2)\|_2^2 \leq \|AF(x_1) - AF(x_2)\|_2^2 \leq (1+\delta)\|F(x_1) - F(x_2)\|_2^2$$

holds for all $x_1, x_2 \in \Sigma_k$, where $F$ is injective for all $x \in \Sigma_k$.

(b) When $F$ is not injective for all $x \in \Sigma_k$, there must exist two distinct $k$-sparse signals $x_1, x_2$ such that $F(x_1) = F(x_2) = z$. Because the image of $z$ under the mapping $L_A$ or matrix $A$ is unique, we have that $AF(x_1) = AF(x_2)$. This means that $\Phi$ is not an injection for all $k$-sparse signals.

In short, the conditions satisfied by $A$ and $F$ all serve to ensure that the composite mappings $\Phi$ are injective for all $k$-sparse signals. When $\Phi$ is not viewed as a composite mapping but a single one, we can study the requirements for $\Phi$ of sparse recovery as well, by other methods. For example, to approximate $\Phi(x)$ using an affine Taylor series type approximation and derive the conditions enabling the IHT algorithm to converge [2].

In the nonlinear CS examples, $F$ is typically not an injection. It makes that, in (1.3), the $k$-sparse solutions consistent with the nonlinear measurements $y$ are possibly more than one, and that the true solution is not necessarily the sparsest one in them. Therefore, when solving (1.3) or (1.4), we may consider adding extra constraints for further reduction of the error with the true solution or improvement of recovery algorithm performance. For example, the unit length constraint in 1-bit CS [3, 15, 22] and the $\ell_1$ norm of the true signal in CSPR [16].

**3.5.2 Pointwise linearization method**

Next, we will introduce a special linearization method, which is referred to as pointwise linearization method. It can transform the nonlinear CS based on composite mappings, pointwise, into linear CS and attain the goal of accurate recovery of all given sparse signals.

Theorem 3.1 below shows that, when the nonlinear mappings $F$ in the composite mappings $\Phi$ can be turned into certain types of pointwise linear mappings, we can transform the composite nonlinear CS pointwise into linear CS, thereby changing the sparse recovery problem in the nonlinear setting into that of the linear setting. Moreover, even if multiple distinct sparse signals have the same nonlinear projection, they can still be exactly recovered via solving their respective linear CS optimization problems (1.1) that are different from each other.

**Theorem 3.1.** *For a given $x \in \Sigma_k$, assume that its nonlinear measurements under a composite mapping $\Phi$ is $z = \Phi(x)$, where $\Phi(x) = F(Ax)$ or $\Phi(x) = AF(x)$. If the sensing matrix A satisfies the RIP of order 2k and the nonlinear mapping F satisfies Definition 3.2, one can always recover $x$ exactly by solving some $\ell_0$ minimization problem (1.1) of linear CS.*

*Proof.* Let $M_I$ be an invertible matrix, and $M_D$ be a matrix produced by any permutation of the rows or columns of an invertible diagonal matrix. Because $F$ satisfies Definition 3.2, for the given $x$, there must exists some matrix $M_I$ or $M_D$ such that $F(Ax) = M_I Ax$ or $AF(x) = AM_D x$. So the nonlinear measurements $z = F(Ax)$ or $z = AF(x)$ are turned, pointwise, into the equivalent linear measurements $z = M_I Ax$ or $z = AM_D x$. In other words, $z$ can be viewed as the projection of the $k$-sparse signal $x$ under the linear composite mapping $M_I A$ or $AM_D$. Moreover, from Theorem 2.3 and that $A$ satisfies the RIP (symmetric or asymmetric) of order $2k$, we know that $M_I A$ and $AM_D$ all satisfy the asymmetric RIP of order $2k$, and their products with $\lambda = \sqrt{2/(\beta + \alpha)}$, i.e., $\lambda M_I A$ and $\lambda AM_D$, all satisfy the symmetric RIP of order $2k$ with $\delta_{2k} = (\beta - \alpha)/(\beta + \alpha) < 1$. We already know that, when $\delta_{2k} < 1$, $\ell_0$ minimization problem (1.1) of linear CS has a unique $k$-sparse solution. Hence, we can exactly recover $x$ by solving the following $\ell_0$ minimization problem:

$$\hat{x} = \arg\min_{u} \|u\|_0 \quad \text{subject to} \quad \lambda M_I Au = \lambda z \quad (\text{or} \quad \lambda AM_D u = \lambda z). \qquad \square$$

In Theorem 3.1, when $\Phi$ is not injective for all $k$-sparse signals, apparently it will result in the situation that multiple distinct $k$-sparse signals are consistent with the same nonlinear measurements $z$. Nonetheless, the matrices $M_I$ or $M_D$ corresponding to these sparse signals certainly differ from each other, and so do their products with $A$, i.e., $M_I A$ or $AM_D$. If not so, it would lead to that two distinct $k$-sparse signals have the same projection under the same product matrix $M_I A$ or $AM_D$, which contradicts its RIP of order $2k$. It is the RIP of these product matrices that guarantees their difference from each other in this situation, as well as the exact recovery of their corresponding sparse signals. Therefore, all of the different sparse signals consistent to $z$ are exactly recovered by solving their respective different $\ell_0$ minimization problems of linear CS.

Theorem 3.1 indicates that, under certain conditions, we can replace the nonlinear mapping $F$ pointwise with a linear mapping $M_I$ or $M_D$, and thus transform the nonlinear composite mapping $\Phi$ pointwise into a linear composite mapping $M_I A$ or $AM_D$, which is a product matrix able to guarantee the accurate reconstruction of its corresponding sparse signal $x$. This is a special pointwise linearization method and it can turn the signal recovery problem of nonlinear CS into that of linear CS.

However, determining $M_I$ and $M_D$ is obviously related to the given sparse signal $x$ (true signal itself or its sparse representation), and generally speaking, it needs to know $x$ or some a priori information unique to $x$. In some applications with true signals known in advance, such as image compression, this pointwise linearization method is applicable.

## 4 Experiments

We will use several examples to demonstrate the pointwise linearization method proposed in this paper. In the composite mapping $\Phi = F \circ L_A$, we take the following three nonlinear functions as $F$. $F_1(x)$ has such a definition that its value is a $m$-dimensional nonzero random vector if $x \neq 0$, and a zero vector if $x = 0$. $F_2(x)$ and $F_3(x)$ are defined respectively as the element-wise absolute value function and signum function. In the composite mapping $\Phi = L_A \circ F$, let $F$ be the following two nonlinear functions. $F_4(x)$ and $F_5(x)$ are defined respectively as the element-wise sine function (restricting the elements of $x$ to $-\pi \leq x_i \leq \pi$) and square function. The function $F_1$ satisfies the requirement of the second type of pointwise linearization, and can be replaced by an invertible matrix $M_I$ at each point. The functions $F_2$ - $F_5$ all satisfy the requirement of the third type of pointwise linearization, and can be replaced by an invertible diagonal matrix $M_D$ at each point. All of the matrices $M_I$ and $M_D$ can be constructed or computed according to Section 3.4.

In all of the experiments, we solve the signal recovery problems using the $l_1$-Magic toolbox. In Figure 4.1, (a) shows that the random sensing matrix $A$, which is chosen according to a Gaussian distribution, has the capability to accurately recover the $k$-sparse signals with $k = 25$; (b) - (f) show that, using the recovery algorithm of linear CS, we can fulfill the accurate reconstruction of given $k$-sparse signals from nonlinear measurements after the nonlinear composite mapping $\Phi$ is replaced by an equivalent linear composite mapping $M_I A$ or $A M_D$ at each point (i.e., applying the pointwise linearization method).

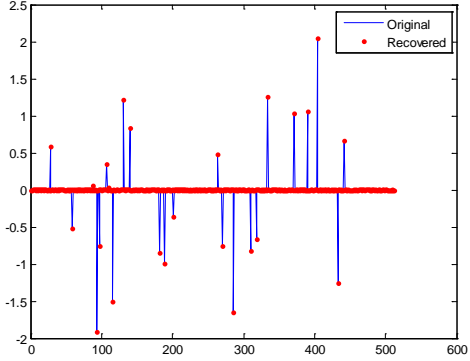

(a)

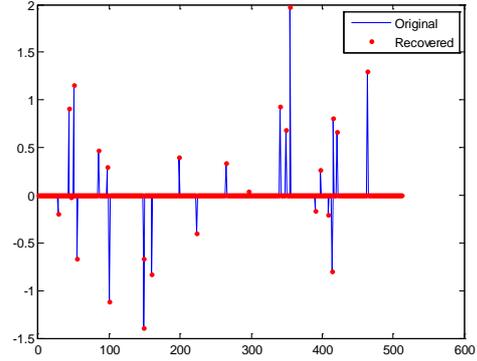

(b)

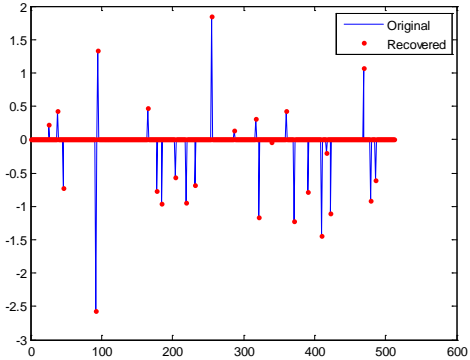

(c)

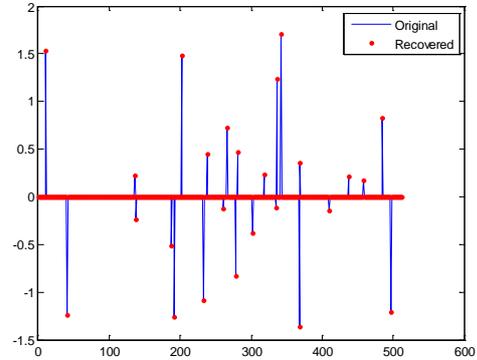

(d)

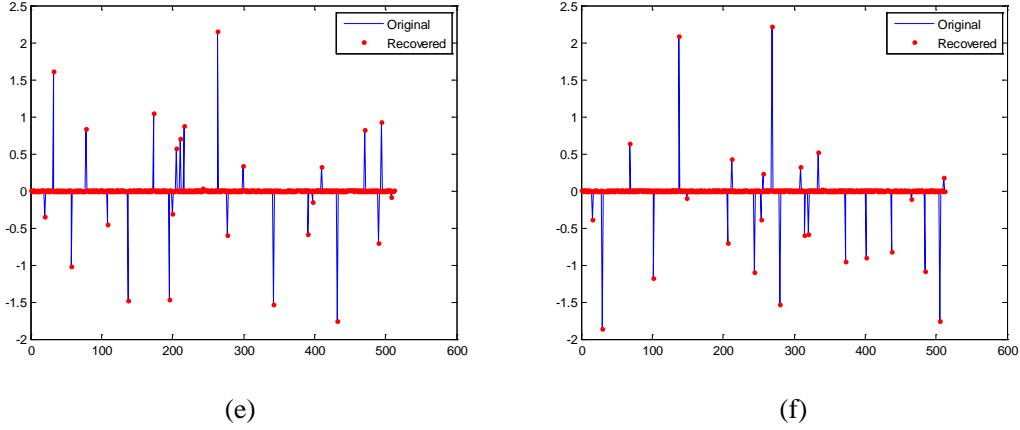

| (e) | (f) |

**Figure 4.1.** The results of recovering a randomly given $x \in \Sigma_k$ from the measurements $Ax$, $F_1(Ax)$, $F_2(Ax)$, $F_3(Ax)$, $AF_4(x)$ and $AF_5(x)$ respectively. (a) shows that the random sensing matrix $A$ satisfies the RIP of order $2k$ and can guarantee the accurate reconstruction of the $k$-sparse signal $x$. (b) - (f) correspond to nonlinear composite mappings $F_1(Ax)$, $F_2(Ax)$, $F_3(Ax)$, $AF_4(x)$ and $AF_5(x)$ respectively. We apply the pointwise linearization method to replace the former three mappings with linear composite mappings of the form $M_I A$ and the latter two with ones of the form $AM_D$. So (b) - (f) represent the results of recovering the $k$-sparse signals $x$ in way of linear CS with these matrix products $M_I A$ or $AM_D$ as sensing matrices.

Note that the quantization error of measurements is generally treated as bounded noise. Therefore, all experiments should be regarded as recovering exactly sparse signals in noisy settings. In these situations, the reconstruction error of an exactly sparse signal will be affected by the RIP constant. The value of the RIP constant is typically altered if one multiplies the sensing matrix $A$ by $M_I$ or $M_D$. Hence, we need to choose $M_I$ or $M_D$ (if multiple exist) appropriately so as to control the size of the RIP constant and thereby of the reconstruction error, for example, in experiment (e) and (f).

## 5 Conclusions

This paper studies the sparse recovery of the nonlinear CS based on composite mappings from the perspective of mapping decomposition, by analyzing the practical examples of nonlinear CS. Generally speaking, for a nonlinear composite mapping $\Phi$, the requirements of exact recovery are quite harsh, which make it impossible to accurately recover all sparse signals in most of the nonlinear CS examples. In light of the invariance of sparse recovery properties under elementary transformations of matrices, we propose a pointwise linearization way which can transform the composite nonlinear CS, at each point, into a composite linear CS, and make us able to accurately recover all sparse signals from nonlinear measurements. The linearization method may bring about an algorithm framework for the composite nonlinear CS, which permits the use of the recovery algorithms belonging to linear CS to reconstruct signals from nonlinear measurements.

However, the nonlinear CS based on composite mappings, proposed in this paper, still has some problems to be explored later. Take for examples the following two problems. Can we recover any given sparse signal by solving a $\ell_1$ minimization problem after pointwise linearization? And is it possible for us to employ the pointwise linearization method in practice?

Generally, if we multiply a sensing matrix $A$ by $M_I$ or $M_D$, the value of its RIP constant $\delta$ will change. We know from above that the solution to the $\ell_1$ minimization problem is that of the $\ell_0$ minimization problem only when $\delta_{2k} < \sqrt{2} - 1$. The $\ell_1$ minimization algorithm used in the experiments is able to reconstruct some random given sparse signal almost accurately, which means that the RIP constant of the corresponding $M_I A$ or $AM_D$ satisfies $\delta_{2k} < \sqrt{2} - 1$ with high probability. So, is it possible to choose the appropriate $A$ and $F$ such that, for all given sparse signals, we can transform $\Phi$ by pointwise linearization into some $M_I A$ or $AM_D$ with $\delta_{2k} < \sqrt{2} - 1$?

If possible, it means that we can also solve a $\ell_1$ minimization problem to recover any given sparse signal exactly in Theorem 3.1. This is a research area we will consider in the future.

When multiple distinct sparse signals have overlapped projections under the nonlinear composite mapping $\Phi$, naturally one cannot recover these signals exactly based solely on the nonlinear measurements by any method whatsoever. Hence, it is no surprise for us that the application of the pointwise linearization method typically need true signals themselves to determine the matrix $M_I$ or $M_D$, although the method can reconstruct all sparse signals accurately. This severely restricts its practical application. For the present, one of its possible application area is image compression. Specifically, after a known image *x* is compressed by a sensing matrix *A*, if the compressed data *Ax* is damaged by nonlinear interference *F* in the process of network transmission we could employ the method to model *F* as a linear matrix mapping $M_I$. Then, the data recipient could recover the original image from the damaged data in way of linear CS, by taking $M_I A$ as a sensing matrix, in the case that the image *x* needs to be transmitted repeatedly (for example, the recipient device wants to release its memory space regularly). Obviously, because the method can make the best of the currently mature algorithms for sparse recovery, how to get it a better application becomes one of the subjects for us to research in the future.